# Tunable discontinuous shear thickening in capillary flow of MR suspensions

Georges Bossis,[1] Yan Grasselli,[2] Alain Ciffreo[1] and Olga Volkova[1]

[1]*Université University Côte d'Azur, CNRS UMR 7010, Institute of Physics of Nice, Parc Valrose 06108 Nice, France*
[2]*SKEMA Bachelor – 60 rue Dostoievski – BP085 – 06902 Sophia Antipolis, France*

Abstract

Very concentrated suspensions of iron particles in water or ethylene glycol can be obtained thanks to the use of superplasticizer molecules used in cement industry. At high volume fractions, these suspensions show a discontinuous shear thickening which was thoroughly characterized in rotational geometries. We will show that the jamming transition is also present in a capillary flow, and that it manifests through the formation of a non-consolidated porous medium at the constriction between the barrel and the capillary. In suspension of iron particles, the dynamics of formation of this porous medium, and so the pressure, can be controlled by a low magnetic field and is reversible for a constant volume flow rate, opening potential new applications in the domain of dampers and force control devices.

# Introduction

Discontinuous shear thickening (DST), also called extreme shear thickening is a phenomenon which occurs in very concentrated suspensions and which is characterized by a sudden jump of stress if we impose the shear rate or a decrease of the shear rate if we increase the stress.

In a pioneering work (Hoffman, 1982, 1972) on monodisperse suspensions, this sudden jump of viscosity was attributed to an order-disorder transition where the organization of the particles in layers flowing over each other was suddenly broken when the stress was increased above a certain level. This sudden increase in viscosity was observed both in cone-plate geometry and in capillary flow (Laun et al., 1991) although the critical shear rate was two times higher in capillary flow. More recently there was a large renew of interest for discontinuous shear thickening, see for instance reviews (Brown and Jaeger, 2014; Denn et al., 2018) and there is now a consensus that this jamming transition is due to the sudden formation of a percolated network of frictional contacts which can behave like a solid skeleton able to support a high stress with low deformation rate. These frictional contacts arise when the imposed stress on a given pair of particle overcomes the repulsive force which, in stabilized suspensions, prevents the aggregation. Then the criteria for the onset of the DST is that the compressive force $F \propto \pi a^2 \sigma$

(with a the radius of the particles and σ the applied stress) becomes larger than the repulsive one given either by an ionic layer in polar solvent (Franks et al., 2000; Singh et al., 2019) or by a coating layer made of small polymer adsorbed or attached to the surface of the particles. In this last case the transition occurs when the polymer desorbs from the surface on the effect of the applied stress (Bossis et al., 2017a; Klein, 2013; Raviv et al., 2001). This percolation mechanism was also evidenced by numerical simulation (Mari et al., 2014). On the other hand, if with the help of a magnetic or of an electric field, we increase the compressive force between the particles, then we expect that it can trigger the onset of the DST transition. That is what we have shown previously in plate-plate geometry with a suspension of carbonyl iron particles at high volume fraction (Φ>0.6) thanks to the use of a superplasticizer molecule from cement industry (Bossis et al., 2016). In many practical applications of magnetorheological (MR) suspensions, the flow of the fluid occurs between fixed walls, either in a capillary tube or in a restricted space between two planes or two cylinders. We previously presented some preliminary results (Bossis et al., 2019b) of the pressure versus flow rate curve for a MR suspension in a capillary rheometer (Malvern RH7) where the jamming transition was clearly identified. Nevertheless it was not possible to install some coils on this apparatus in order to study the effect of the magnetic field on the jamming transition. Also, in this vertical capillary rheometer we came with the problem of sedimentation and change of density in the bottom of the barrel in front of the entrance of the capillary. To overcome these difficulties we have developed a capillary rheometer based on a syringe pump; we describe the apparatus in the section "materials and methods". In the second part we compare the jamming obtained in plate-plate rheometry to the one obtained with our home made capillary rheometer on the same suspensions for 3 different volume fractions: 61%, 62%, 63% at zero external field and we interpret the results with the help of the measurement of the resistance of the suspension. In the last part we present the results obtained in the presence of the magnetic field and again compare them to the one obtained in plate-plate geometry for the same internal magnetic fields.

# Materials and methods

The homemade capillary rheometer is based on a syringe pump (Nemesys medium syringe pump 1000N) which is shown in Fig.1. The syringe itself has been replaced by a cylinder of internal diameter 25mm and surrounded by a coil. On the left extremity of the piston we have placed an annular force sensor (Mesurex FC1000N8-25A, range 1000N) allowing to fix the piston to the pusher by a screw passing through this annular sensor. The sealing between the cylinder and the head of the piston is made by an O'ring The other extremity of the cylinder is closed by a flange with a capillary hole of diameter 2.2mm surrounded by 4 electrodes (Fig.2). The diaphragm of a miniature pressure sensor (Mesurex PRM5C20B4V, range 0-20bar) reaches the surface on the rim of the inlet cone. The pressure was both obtained from the force sensor and the pressure sensor. Before the jamming transition the indication of the pressure sensor was chosen because it does not integrate the friction force of the O'ring around the head of the piston which can go up to 50N. On the contrary above the transition, the formation of a solid skeleton of particles can support the pressure and partly screens the surface of the sensor, so above the transition, at high pressure we choose the indication of the force sensor. The length of the capillary was 2cm in the bulk of the flange and 6.5cm or 10.6cm when extended with a plastic tube of the same internal diameter fixed with the flangeless system Delrin (1/8").

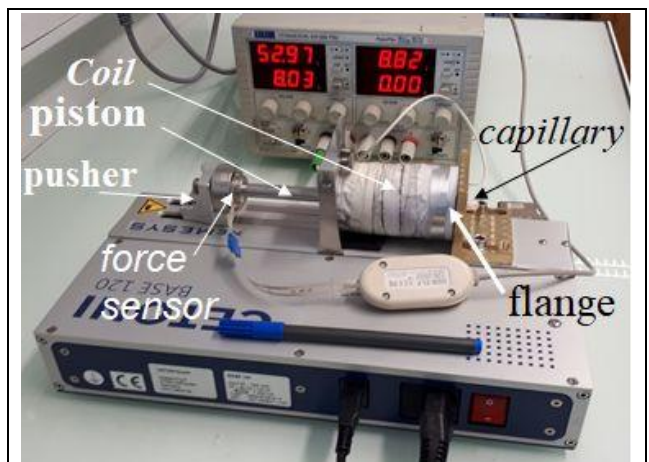

**Figure 1.** Photo of the home made capillary rheometer

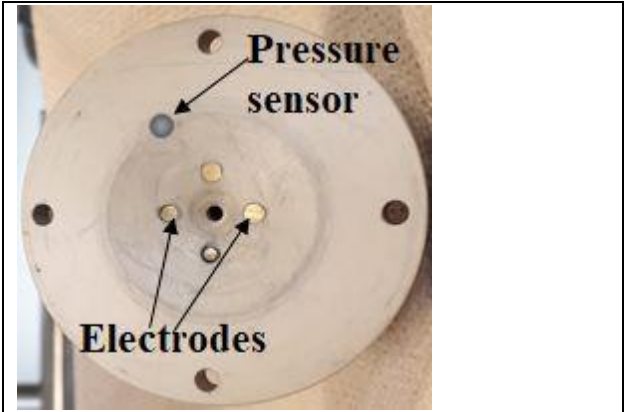

**Figure 2**. Photo of the flange closing the barrel

The angle of the cone was 13° and its diameter 20mm. This flange was pressed by 4 screws on an O'ring placed in a groove at the extremity of the barrel. The MR fluid was made from 99% purity carbonyl iron particles ( supplier: Gongyi City Yalv Material Co.,Ltd). The suspending fluid was a mixture of water (15%) with ethylene glycol with a viscosity µ=0.011Pa.s and the superplasticizer molecule was Optima 100 from Chryso.

## Test of the capillary rheometer

In order to test the device in the same conditions as with the MR fluid, but without the problems related with sedimentation of the particles or yield stress ,we have used an honey of viscosity in the range of 10 Pa.s ; its viscosity ,$\eta_f$, measured at 27°C on the rheometer MCR502, Anton Paar was varying from 8.5Pa.s at 115 $s^{-1}$ to 4.7 Pa.s at 2300 $s^{-1}$. The temperature of 27°C was the room temperature and we checked that, at the output of the capillary, the temperature did not raise more than 2°C at the highest shear rate. We took a length of the capillary L=10.6cm.

In the Fig.3 we have plotted the experimental pressure against the theoretical one for 6 different flow rates Q. For a Newtonian fluid we have respectively for the shear rate at the wall and the pressure drop:

$\dot\gamma = 4Q/(\pi.R^3)$ and $P_{th} = 2L\,\eta_f\dot\gamma/R$   (1)

Where R=1.1mm is the radius of the capillary. Note that with this radius a shear rate of $10s^{-1}$ corresponds to a flow rate of 10.5µl/s.

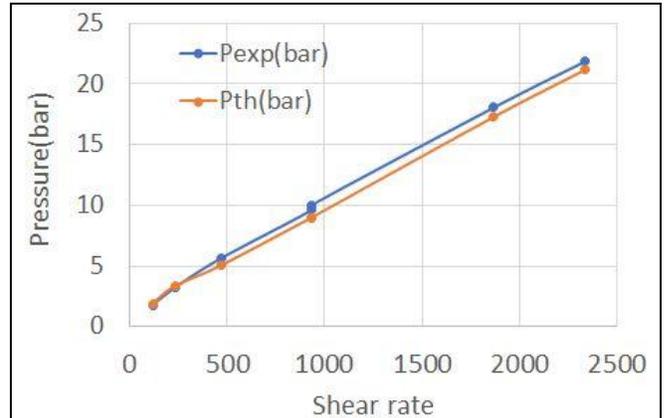

**Figure 3.** Measured pressure drop $P_{exp}$ versus the theoretical one $P_{th}$ obtained from the measurement of the viscosity

The agreement is good at the lowest shear rate; there is a small discrepancy at higher shear rates but most of our measurements will be at shear rates below 200 $s^{-1}$.It is also interesting to test the dynamical response to a jump of shear rate; it is reported in Fig.4

We see that if we impose an abrupt jump of flow rate the pressure follows but with a first shift of 0.1s and an overall delay of about 0.4s before to reach the plateau. Also, when the flow is turned off, the beginning of the drop of pressure is shifted by 0.1s and after a sharp drop there is a slower one of about 1s before returning to zero.

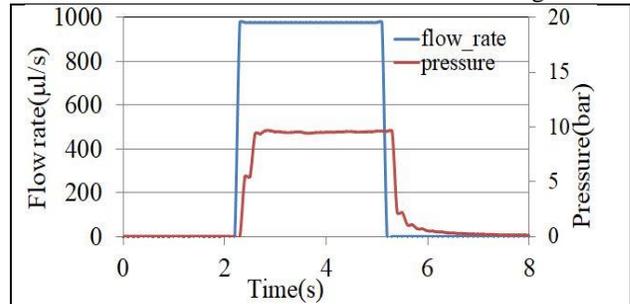

**Figure 4.** Response of the pressure to a jump of flow rate

The origin of the shift of about 0.1s between the response of the pressure and the one of the flow could be due to the delay between the command of the motor of the syringe and the real start of the flow, or also from some hysteresis of the diaphragm of the pressure sensor; in any event it is small compared to the time scale that we shall observe in the following. The longer tail when the flow is stopped comes from the exponential decompression of the liquid with a time constant proportional to the compressibility modulus and inversely proportional to the viscosity; it will always be present.

## Jamming in the absence of field

We begin by a presentation of the jamming phenomenon in a capillary in the absence of a magnetic field. The phenomenon of jamming in dense suspensions has been reported a long time ago; see for instance (Laun et al., 1991). More recently, in an experiment where the suspension was sucked inside a capillary it was shown, using brightfield microscopy, that there was the formation of a jammed plug of particles at the entrance of the capillary and some self filtration through this porous plug which was moving at a smaller velocity than the suspending fluid (Haw, 2004). Similar conclusions were drawn with the help of confocal microscopy in a pressure driven flow (Isa, 2008). By direct visualization of the flow of an alumina suspension in a slit capillary (Han and Ahn, 2013), the authors found the presence of an important shear banding and of a velocity profile with two inflections points which is very different from the parabolic one of a Newtonian fluid. In the suspensions of iron particles it is not possible to have some information on the velocity profile by microscopy, but as the particles are conductive, we have placed electrodes (cf Fig.2) around the entrance of the capillary in order to follow the change of resistance and thus to get an information on the compaction of particles at the entrance of the capillary. Our aim here is first to catch the jamming phenomenon and then to compare the critical shear rate corresponding to the rise of the pressure with the one obtained on a conventional rheometer in torsional flow between two plates. The stress-shear curve of the suspension of iron particles at three different volume fractions is presented in Fig.5. The DST transition is well defined and, at the highest volume fraction, is characterized by a sudden decrease of the shear rate. In both cases the shear rate fluctuates strongly after the transition. Whereas, for the two higher fractions, the behavior below the transition can be correctly described by a Bingham law, this is not the case for the lower volume fraction, where we have a shear thickening part before the transition. Also the critical shear stress is more higher at 61% than for the two other volume fractions. Actually it means that the volume fraction $\Phi= 61\%$, for this given composition of the suspension, is close to the volume fraction below which the DST transition disappears to give place to the regime of continuous shear thickening. This volume fraction is the smallest one delimiting the zone between the shear jammed and unjammed situations in a $(\tau, \phi)$ phase diagram (Bi et al., 2011). In Table 1 are reported the values of the critical stress and critical shear rate, together with the parameters of a Bingham law: $\tau = \tau_y + \eta_{pl}\dot{\gamma}$. Note that for a non Newtonian fluid the experimental curve $\tau_e(\dot{\gamma})$ obtained with a plate-plate geometry should be corrected by using the Mooney-Rabinovitch equation: $\tau = \frac{\tau_e}{4}\left[3 + \frac{\dot{\gamma}}{\tau_e}\frac{d\tau_e}{d\dot{\gamma}}\right]$. If $\tau_e$ follows a Bingham law: $\tau_e = \tau_{ya} + \eta_{pl}\dot{\gamma}$ then we get: $\tau = \frac{3}{4}\tau_{ya} + \eta_{pl}\dot{\gamma}$ so the true yield stress is $\tau_y = 0.75\tau_{ya}$ with $\tau_{ya}$ the apparent yield stress. On the other hand the plastic viscosity $\eta_{pl}$ remains equal to the slope of the experimental curve in plate-plate geometry. For $\Phi=62\%$ and $\Phi=63\%$ we took it at the critical point:

$$\eta_{pl} = (\tau_{cr} - \tau_{ya})/\dot{\gamma}_{cr} \qquad (2)$$

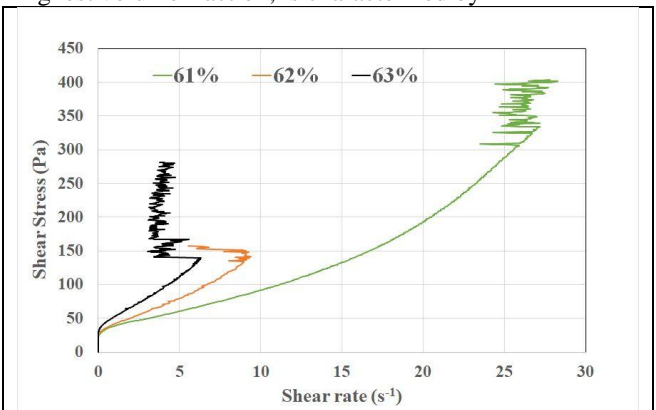

**Figure 5.** Rheogram of carbonyl iron suspensions in plate-plate geometry

For $\Phi=61\%$, where there is a quite strong shear thickening before jamming (cf Fig.(5)), we took Eq.(2) at half the critical stress ($\tau=160$Pa, $\dot{\gamma} = 17 s^{-1}$) where a Bingham approximation is more justified.

|  | $\tau_{ya}$ (Pa) | $\tau_y$ (Pa) | $\tau_{cr}$ (Pa) | $\dot{\gamma}_{cr}$ ($s^{-1}$) | $\eta_{pl}$ (Pa.s) |
|---|---|---|---|---|---|
| $\Phi=61\%$ | 22.5 | 16.9 | 324 | **26.6** | 8 |
| $\Phi=62\%$ | 24 | 18 | 131 | **8.8** | 12 |
| $\Phi=63\%$ | 28.5 | 21.4 | 138 | **6.3** | 17 |

Table 1: Experimental data from the rheograms of Fig.5

Now we are going to see how the jamming transition occurs in a capillary flow for these same suspensions. The experiments used to determine the critical stress and the critical shear rate were made with a ramp of flow rate. We have shown in Fig.6 a ramp of flow rate versus time (left axis) and the resulting pressure (right axis) for a suspension at $\Phi=61\%$. The red curve is the prediction for the flow of a Bingham fluid in a cylindrical capillary. The Buckingham equation (Sisko, 1958) relates the flow rate Q to the yield stress and to the wall shear rate, $\tau_w$:

$$Q = \frac{\pi}{4} \cdot \frac{R^3}{\eta}\left[\tau_w - \frac{4}{3}\tau_y + \frac{\tau_y^4}{3\tau_w^3}\right] \text{ with } \tau_w = \frac{P.R}{2.L} \qquad (3)$$

Solving numerically this equation gives the relation between the pressure drop P and the flow rate Q. Even if the Bingham law is a rough approximation at $\Phi=61\%$, it gives the right order of magnitude of the pressure in the absence of jamming.

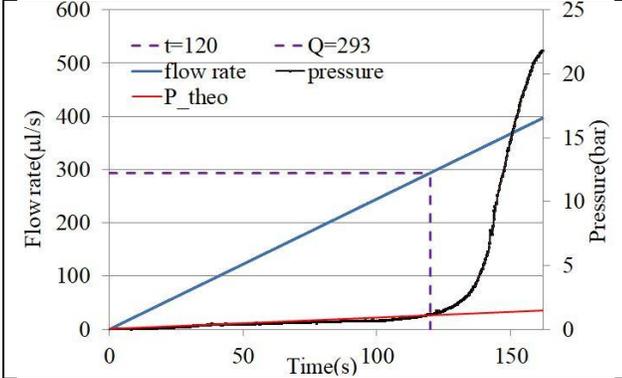

**Figure 6.** Flow rate (in blue) and pressure (in black) versus time for Φ=61%. The red line is the theoretical pressure of a Bingham fluid (Eq.(3)) with the viscosity and yield stress reported in Table 1

The point where the pressure leaves this monotonic behavior to raise sharply defines the critical pressure, $P_c$ and the critical flow rate (dotted line in Fig.6).
From $P_c$ we have the critical stress on the wall, $\tau_{cw}$, (Eq.(3)) and so the critical shear rate in the capillary flow:
$$\dot{\gamma}_{cr(cap)} = (\tau_{cw} - \tau_y)/\eta \qquad (4)$$
The comparison between the critical shear rate and the critical stress obtained respectively in plate-plate geometry and in cylindrical capillary is reported in Table 2 where L is the length of the capillary

| Φ, L | 61%, 2cm | 62%, 6.5cm | 63%, 2cm |
|---|---|---|---|
| $\dot{\gamma}_{cr(plate)}(s^{-1})$ | 26.6 | 8.8 | 6.3 |
| $Q_{cr}(\mu l/s)$ | 293 | 41.5 | 31 |
| $\dot{\gamma}_{cr(cap)}(s^{-1})$ | 281 | 40.2 | 30 |

Table 2: Comparison of the critical shear rates in plate plate geometry and in a cylindrical capillary

The longer capillary at 62% was used in order to check that a change of length did not change the pressure drop per unit length. It appears that for the three volume fractions the critical shear rate calculated in capillary geometry is several times larger than the one obtained in plate-plate geometry. The ratio of the critical stress minus the yield stress will be the same (cf Eqs. (3) and (4)). For a given volume fraction the jamming is triggered by the fact that the applied stress overcomes the repulsive stress provided by the coating layer, so there is no reason that in the capillary the critical stress should be several times the one in plate-plate geometry. One hypothesis is that the jamming does not take place in the capillary, but in the transition zone between the barrel and the entrance of the capillary where the shear rate and so the stress is much lower. In this case the percolating network of particles would form upstream of the capillary and would be convected to the entrance of the capillary, forming a transient plug in front of the capillary. This explanation is well supported by the evolution of the resistance between two diametrically opposed electrodes (cf Fig.2) as is shown in Fig.(7) where is plotted both the evolution of the pressure and of the resistance during a ramp of flow rate at Φ=63%

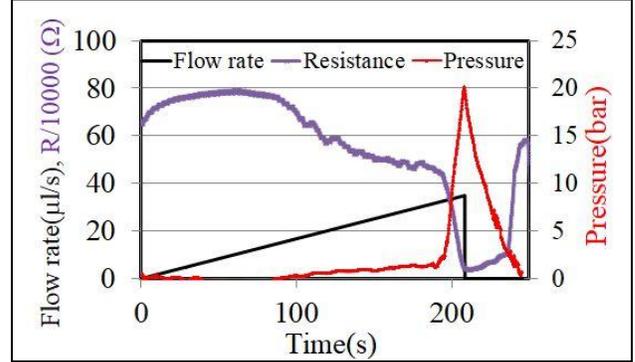

**Figure 7.** Evolution of the resistance(in purple) and pressure(in red) during a ramp of flow rate. Φ =63%

In this figure we can see that the sharp raise of pressure due to jamming goes with a strong decrease of the resistance from 440kΩ to 40kΩ which corresponds to the formation of a layer with an improved conductivity between particles due to their frictional contact. When the pressure reaches 20 bar, the security of the syringe pump stops the flow and the pressure relaxes during the decompression of the cell. During this decompression the resistance increases first slowly and then jump to a large value at the end of decompression, meaning that the interparticle contacts have disappeared. A remaining question is why the transition does not appear first in the capillary when the wall shear rate reaches the critical one. The answer is very likely that this transition is prevented by the formation of a plug flow due to strong particle migration from high shear rate close to the wall towards the center of the capillary (Fall et al., 2010; Mills and Snabre, 1995). Another point to note is that the resistance begins to decrease after the time t=100s and at the same time the pressure gently increases. Contrary to the case at 61% described in Fig. (6), this increase of pressure is quite higher than the one predicted from the yield stress and viscosity given in Table1. For instance at t=190s corresponding to a flow rate of 31μl/s the pressure is equal to 1.23bar but the theoretical one is only 0.19 bar. This correlation between the increase of pressure and the decrease of the resistance at the entrance of the capillary indicates that before the jamming transition where the particles come into frictional contacts, we can have some local increase of the volume fraction inside a layer at the entrance of the capillary, which partially filtrates the suspension and provokes this slight increase of pressure. Actually this phenomenon of filtration is well known in the extrusion of pastes and is described in many papers both experimentally (O'Neill et al., 2015; Rough et al., 2000) theoretically (Khelifi et al., 2013) and with the help of numerical simulations (Ness et al., 2017; Patel et al., 2017).

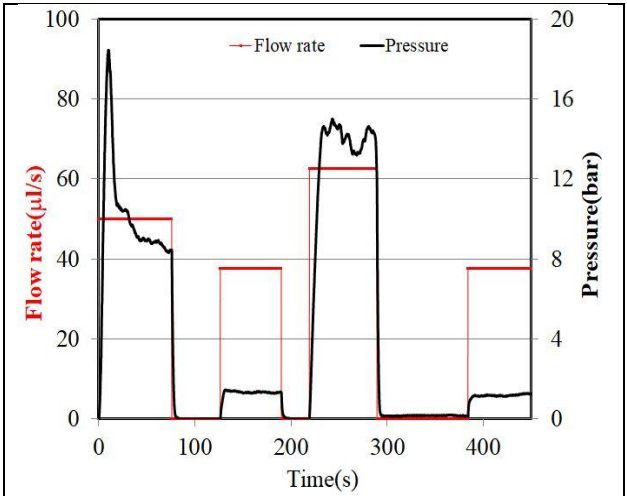

**Figure 8.** Pressure response (in black) to jumps of flow rates (in red); Φ=62%

This filtration process is predominant when the speed of the plunger is slow enough to let time to the plug to form before to be expelled and it is the reason why we see this first increase of pressure at Φ=63% and not at 61% where the flow rate is 10 times larger. Although connected, the jamming transition must be distinguished from the process of densification at the entrance of the die, because it happens suddenly above a certain stress and it is reversible when the stress goes down, as can be seen in Fig.8 for a suspension at Φ= 62%

The critical flow rate here is 41.5 μl/s and we see that passing from 50μl/s to a little bit below 40 we recover the low pressure state and still again after a jump at 60μl/s. Now we are going to see that this kind of jump from the normal state to the jamming state can also be controlled by the application of a magnetic field.

# Jamming controlled by a magnetic field

As was shown in previous papers, the application of the magnetic field can induce the friction between the surfaces of the particles and provokes a jump of stress, which is moderate at low volume fraction (Jiang et al., 2015), but which can reach 150kPa even for fields as low as 8kA/m (Bossis et al., 2017b). This was obtained in conventional rheometry either in a plate-plate geometry or in cylindrical Couette geometry. Our aim here is to see if this jump of stress can also be observed in the capillary geometry. As described in Fig.1 the syringe barrel is surrounded by a coil which allows to impose a magnetic field on the entrance of the capillary. The field that we are going to report here was measured just at the entrance of the capillary. Again we want to compare the effect of the field in plate-plate geometry to the one obtained with the capillary. So we have measured for the three volume fractions the shear stress versus shear rates for different magnetic field in the plate-plate geometry for the same internal field as in capillary flow. For the measurement in plate-plate geometry we have multiplied the field imposed in capillary flow by the permeability of the suspension at low field in order to take into account the demagnetizing field. The magnetic permeability of the suspensions can be deduced from a fit of the initial permeability versus the volume fraction (De Vicente et al., 2002) which is also in good agreement with the values taken from the measurement at Φ=0.64 (Bossis et al., 2019a). The evolution of the permeability with the volume fraction can be represented by:

$\mu(\Phi) = 2.15 + 0.0136\Phi + 0.0008\Phi^2$

It changes from 5.95 for Φ=0.61 to 6.07 for Φ=0.62 and 6.18 for Φ=0.63. We present the rheogram for one volume fraction : Φ= 63% in Fig.9. We see that the jamming transition is maintained in the presence of a magnetic field but with a critical shear rate which decreases quickly with the amplitude of the magnetic field.

In the Table 3, we have reported the yield stress, the plastic viscosity, calculated as in Table 1, and the critical values of stress an shear rate versus the internal field for the three volume fractions.

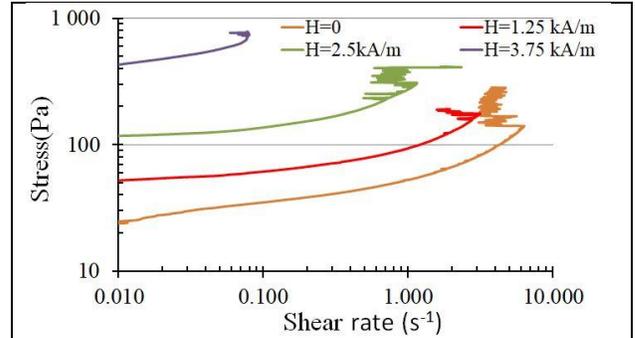

**Figure 9.** Rheogram for different internal magnetic fields; Φ=0.63

| H(kA/m) | $\tau_y$ (Pa) | $\tau_{cr}$(Pa) | $\dot{\gamma}_{cr}(s^{-1})$ | $\eta_{pl}$(Pa.s) |
|---|---|---|---|---|
| 0 | 16.9 | 324 | 26.6 | 8 |
| 1.25 | 37.5 | 420 | 10.8 | 17.4 |
| 2.5 | 84.8 | 439 | 3.51 | 73.5 |
| 3.75 | 157.5 | 528 | 1.06 | 280 |
| 5 | 442.5 | 860 | 0.11 | 1260 |
| H(kA/m) | $\tau_y$ (Pa) | $\tau_{cr}$(Pa) | $\dot{\gamma}_{cr}(s^{-1})$ | $\eta_{pl}$(Pa.s) |
| 0 | 18 | 131 | 8.8 | 12.2 |
| 1.25 | 46.5 | 198 | 5.0 | 27 |
| 2.5 | 90.0 | 327 | 1.8 | 115 |
| 3.75 | 210.0 | 697 | 0.4 | 1017 |
| 5 | 862.5 | >1600 | <0.04 | 8750 |
| H(kA/m) | $\tau_y$ (Pa) | $\tau_{cr}$(Pa) | $\dot{\gamma}_{cr}(s^{-1})$ | $\eta_{pl}$(Pa.s) |
| 0 | 21.4 | 138 | 6.3 | 17.4 |

| 1.25 | 41.3  | 171 | 3     | 39   |
|------|-------|-----|-------|------|
| 2.5  | 90.0  | 302 | 1.14  | 160  |
| 3.75 | 315.0 | 667 | 0.075 | 3293 |

Table 3  From top to bottom: Φ= 61%, Φ=62%, Φ=63%

In MR fluid at low field the yield stress increases with a power law $\tau_y \propto H^n$ with n≤ 2, (Ginder et al., 1996) but here it is larger than in conventional fluid.  For instance at H=3.75kA/m we have an increase of yield stress compared to the one at zero field of about 300Pa for Φ=0.63, whereas the theory based on finite element calculation of the magnetic forces between spheres predicts a value of 45 Pa   (Bossis et al., 2019a). Also we have noted that the increase of the yield stress with the field is higher than the one  predicted by the $H^2$ law.

Since the yield stress is predicted to be proportional to the number of chains of particles induced by the application of the magnetic field, it should be proportional to the volume fraction, that is to say almost the same for the three volume fractions at the same field. As can be seen from the inspection of the Table 3, this is not at all the case : for instance at H=3.75kA/m it practically doubles between 61% and 63%. All these observations are related to the existence of a supplementary frictional stress induced by the magnetic field which increases much faster than linearly with the volume fraction. We shall develop this point in a future paper; here we are more interested by the comparison which can be done concerning the effect of the magnetic field on jamming in capillary flow.

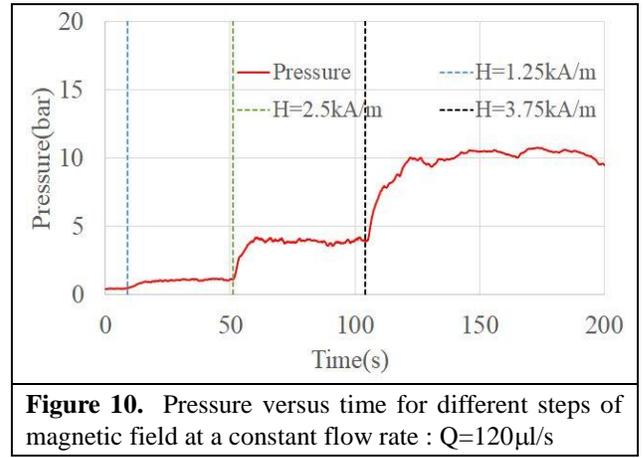

**Figure 10.**  Pressure versus time for different steps of magnetic field at a constant flow rate : Q=120μl/s

We show in Fig. 10 the change of pressure with the magnetic field at a constant flow rate, Q=120μl/s well below the critical one (Qc=293μl/s). The time where the magnetic field is changed corresponds to the dotted vertical line. The next field H=5kA/m was also applied at the same flow rate in an other experiment not shown here
All the results in terms of the pressure jump after the application of the field  are reported in the Table 4 for the volume fractions Φ=61% and Φ=63%. In the two last columns we have compared the theoretical pressure jump to the experimental one. The theoretical pressure is the one corresponding to the flow of a Bingham fluid with the values reported in Table 3 for the yield stress and the plastic viscosity, that is to say in the absence of the jamming transition. At the volume fraction of 61% the experimental pressure is in agreement with the predictions, except at the higher field, so we could conclude that the field does not induce a jamming transition.

| Φ, H | $\dot{\gamma}_{cr}$(s⁻¹) plan-plan | Q_ (μl/s) | $\dot{\gamma}_{cap}$(s⁻¹) | P_theo (bar) | P_exp (bar) |
|---|---|---|---|---|---|
| 61%  (H=1.25kA/m) | 10.8 | 120 | 118 | 0.76±0.08 | 0.6 ± 0.1 |
| 61% (H=2.5 kA/m) | 3.51 | 120 | 117 | 3.16±0.3 | 3.4 ± 0.2 |
| 61% (H=3.75kA/m) | 1.06 | 120 | 117 | 11.9±1.2 | 9.6 ± 0.5 |
| 61% (H=5kA/m) | 0.11 | 120 | 117 | 54±5.6 | 17±1 |
|  |  |  |  |  |  |
| 63%  (H=1.25kA/m) | 3 | 25 | 24 | 0.36±0.04 | 17±1 |
| 63% (H=2.5kA/m) | 1.14 | 25 | 24 | 1.43±0.2 | >20 |
| 63% (H=3.75kA/m) | 0.07 | 25 | 24 | 28.8±0.2 | >20 |

Table 4  Comparison of the theoretical pressure drop (P_theo) in the presence of a Bingham flow in the capillary of length L=2cm ,with the experimental one

and that he change of pressure is just the result of the classical magnetorheological effect. This is not the case at Φ=63% where, for the two lower fields, the experimental pressure is at least one order of magnitude larger than the one of a pure Bingham fluid. In this situation it is clear that the application of the field has triggered the jamming.

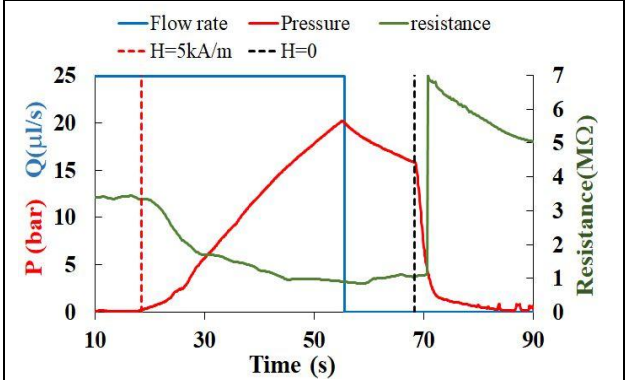

**Figure 11.** Evolution of the pressure and of the resistance versus time after the application of a field H=5kA/m at t=20s which is turned off at t=69s

We have imposed a constant flow rate Q=25μl/s below the critical one (Q=31μl/s). After the application of the field H=5kA/m at the time t=20s we observe a decrease of the resistance from 3MΩ to 1MΩ during the increase of the pressure. This is the same behavior as without field and it indicates the formation of a denser layer of particles with an improved interparticle conductivity around the orifice of the die. When the pressure reaches 20 bar the flow stops, but we have maintained the field. The pressure decreases slowly, indicating that we have a small leakage through this denser layer which form a plug at the entrance of the die. When the field is turned off at t=69s the pressure drops abruptly and during this drop we have also an abrupt increase of the resistance, indicating the destruction of this dense network of particles at the entrance of the die.

Another aspect of this transition which needs to be deepened is its dynamics. Actually looking at Fig.11 we can see that it is quite slow since we attain 20bars in 35s. The evolution of the pressure with time is shown in Fig.12

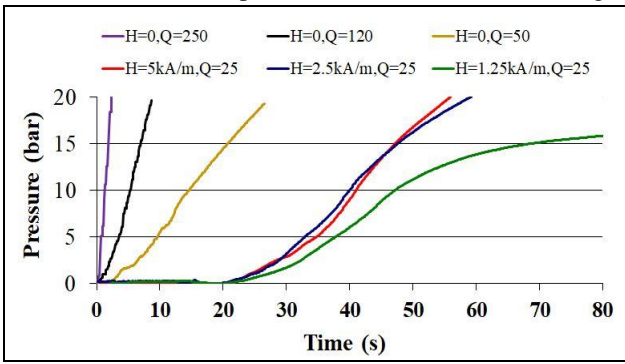

transition. Now as for the case at zero field, one might ask if the transition takes place inside the die or at its entrance. We have measured the change of resistance together with the pressure when we apply a field of 5kA/m at the time t=20s. The result is represented in Fig.11

**Figure 12.** Pressure versus time for different fields and flow rates at volume fraction Φ=63%

The three curves beginning at t=0 correspond to a zero field and to different rectangular pulse of flow rates above the critical one ( $Q_c$=31μl/s at Φ=63%). The limit pressure is always 20 bar. The time needed to reach this value was respectively 26.4, 8.7 and 2.4s for flow rates of 50,120,250 μl/s, so it decreases faster than proportionally to the flow rate. On the other hand we have plotted in the same figure the pressure for three different fields applied at t=20s during a flow rate Q=25 μl/s smaller than the critical one. For the smaller field, H=1.25kA/m, the pressure does not reach the value of 20 bar and the rising time up to a plateau of pressure of 17bar is 120s. At a higher field, H=2.5kA/m, we already have a saturation of the pressure profile which is practically the same as the one at 3.75kA/m (not shown for clarity) and 5kA/m. For these fields the rising time is 38s that is to say longer than at zero field with Q=50μl/s. Since the jamming phenomenon is supposed to come from the percolation of a network of frictional contacts we would expect that it takes place as rapidly as in the plate-plate geometry (in a time which should be a fraction of the inverse of the shear rate, so much less than 1s), but here the phenomenon is much slower. What appears is that this rising time depends essentially of the flow rate. It is clear at zero field, but it is also true in the presence of the field. For instance at H=5kA/m and Q=25μl/s the rising time is 28s but for the same field (at Φ=61%) the rising time decreases to 13s at Q=120μl/s. If we look to the time scale of the change of resistance (cf Fig.11 or Fig.7) it is the same as the one of the pressure, so we can assume that it is the time needed to build this layer of particles in frictional contact in front of the die, which triggers the increase of pressure. It can be informative to estimate the pressure drop in terms of the filtration through a porous layer made by the network of jammed particles. It is given by the Carman-Kozeny equation for a porous medium made of spherical beads of diameter d (Patel et al., 2017):

$$P_{por} = L.\eta_f \frac{Q}{\pi R_b^2}.\frac{1}{K} \quad with \quad K = \frac{\varepsilon^3 d^2}{180(1-\varepsilon)^2}$$
(17)

Where the constant of Kozeny is taken equal to 5 and the specific area of the particles is 6/d ; ε=1−Φ is the porosity $\eta_f$ the viscosity of the suspending fluid ($\eta_f$ =0.011Pa.s) and $R_b$ =1.25cm the radius of the barrel. Taking an average diameter of 2μ for the particles and a dense layer of 10 diameters of particles (L=10d) we obtain, for a flow rate of 25μl/s, $P_{por}$=5.1 bar which is the right order of magnitude. Nevertheless it is not realistic to think that such a pressure could be supported by a layer of 20μ. If it was the case we should have a perfect filtration and a flow of the pure suspending fluid at the exit of the capillary, which is not at all the case. Instead we must think about a dynamic plug which is first forming gradually by adding particles with the flow -causing the dependence of the pressure with the flow rate- and reaching an equilibrium size with an average velocity smaller than the superficial velocity $Q/\pi R^2$ which produces the pressure gradient but does not prevent them to flow inside the capillary. Referring to recent simulations of extrusion of pastes in similar geometry (Patel et al., 2017) liquid migration and change of volume fraction of about 0.5% is expected at low values (for instance 0.002) of $V/R_b=Q/(\pi R_b^3)$ and we are in this case since $V/R_b$= 0.004 for Q=25μl/s. We have tried to measure the decrease of the volume fraction of the extrudate, but the change observed (1-2%) was in the limit of uncertainty of our experimental conditions.

## Conclusion

In this work we have tried to see if it was possible to observe the jamming transition in a cylindrical capillary and, above all, if it was possible to control this transition with a magnetic field, using a suspension of iron particles at high enough volume fraction to obtain this transition. At zero field we observe a clear transition with an increase of pressure more than an order of magnitude above the one which characterizes a Bingham flow of the fluid in the absence of the jamming transition (cf. Fig.6). Nevertheless the critical shear stress and the critical shear rate calculated on the wall of the capillary was between 5 and 10 times the one we have measured on a conventional rheometer in plate-plate geometry. This observation together with the fact that we have highlighted the existence of a resistivity drop on a layer of particles resting on the die entry plane, indicates that the jamming transition occurs in a layer at the entrance of the die and not inside. By applying a low magnetic field, we can trigger an increase of pressure, but it is only at Φ=63% and at low fields that we can attribute the increase of pressure to the jamming transition (Table 4).

Furthermore the rate of increase of pressure depends strongly on the flow rate and in practice it can be a limitation for applications working at low flow rates. The mechanism of formation of a jammed layer of particles at the entrance of the die is certainly complex and deserves further work especially using X ray tomography to get a profile of volume fraction inside the extruder. It would be also useful to be able to measure with a good precision the exit volume fraction continuously during the advance of the piston in order to characterize the filtration if any.

## Acknowledgments

The authors want to thank the CENTRE NATIONAL D'ETUDES SPATIALES (CNES, the French Space Agency) for having supported this research and the company CAD for his help in developing the high torque rheometer